\documentclass[aps,pre,twocolumn,showpacs]{revtex4}
\usepackage{graphics}
\usepackage{epsfig}
\usepackage{amsmath}
\usepackage{amssymb}
\usepackage{bm}
\usepackage{color}

\newcommand{\F }{{\bf F}}

\newcommand{\nominal }{{\bf S}}
\newcommand{\thermo }{{\bf T}}
\newcommand{\cauchy}{{\bm \sigma}}
\newcommand{\lagrangian}{{\bm \eta}}
\newcommand{\linstrain}{{\bm \epsilon}}
\newcommand{\Grad}{\mathrm{Grad}\,}
\newcommand{\grad}{\mathrm{grad}\,}
\newcommand{\Div}{\mathrm{Div}\,}
\newcommand{\divv}{\mathrm{div}\,}

\newcommand{\x }{{\bf x}}
\newcommand{\X }{{\bf X}}

\newcommand{\U}{{\bf u}}

\newcommand{\sig }{{\bm \sigma}}

\newcommand{\srr}{\sigma_{rr}}
\newcommand{\lnR}{\ln{\left(R/R_0\right)}}
\newcommand{\Tr}{\mathrm{Tr}\,}
\newcommand{\T}{\mathrm{T}}
\newcommand{\Qm}{Q_m}
\newcommand{\Qd}{Q_d}
\newcommand{\Qq}{Q_q}

\begin{document}

\title{Nonlinear elastic stress response in granular packings}

\author{Brian P.~Tighe}
\altaffiliation[Current address: ]{\em Instituut-Lorentz, Universiteit Leiden, Postbus 9506, 2300 RA Leiden, The Netherlands}
\affiliation{Department of Physics and Center for Nonlinear and Complex Systems, \\
         Duke University, Durham, NC 27708}
\author{Joshua E.~S.~Socolar}
\affiliation{Department of Physics and Center for Nonlinear and Complex Systems, \\
         Duke University, Durham, NC 27708}

\date{\today}

\begin{abstract}
  We study the nonlinear elastic response of a two-dimensional
  material to a localized boundary force, with the particular goal of
  understanding the differences observed between isotropic granular
  materials and those with hexagonal anisotropy. Corrections to the
  classical Boussinesq result for the stresses in an infinite
  half-space of a linear, isotropic material are developed in a power
  series in inverse distance from the point of application of the
  force.  The breakdown of continuum theory on scales of order of the
  grain size is modeled with phenomenological parameters
  characterizing the strengths of induced multipoles near the point of
  application of the external force.  We find that the data of Geng et
  al.~\cite{geng01} on isotropic and hexagonal packings of
  photoelastic grains can be fit within this framework.  Fitting the
  hexagonal packings requires a choice of elastic coefficients with
  hexagonal anisotropy stronger than that of a simple ball and spring
  model.  For both the isotropic and hexagonal cases, induced dipole
  and quadrupole terms produce propagation of stresses away from the
  vertical direction over short distances.  The scale over which such
  propagation occurs is significantly enhanced by the nonlinearities
  that generate hexagonal anisotropy.
\end{abstract}

\maketitle

\section{Introduction}

The response of a granular medium to a localized boundary force has
been investigated both experimentally and numerically \cite{geng01,
  geng03, serero01, reydellet01, mueggenburg02, spannuth04, head01,
  gg04, gg05, kasahara04, moukarzel04, gland05, ellenbroek05, ellenbroek06,
  ostojic05, ostojic06}.  Experiments have shown that in disordered
packings stress response profiles consist of a single peak that
broadens linearly with depth \cite{geng03, serero01}.  For hexagonal
packings of disks \cite{geng01, geng03} or face-centered cubic
packings of spheres \cite{mueggenburg02, spannuth04}, on the other
hand, the stress response develops multiple peaks that seem to
coincide with propagation along lattice directions.  In two
dimensions, a hexagonal packing is indistinguihable from an isotropic
one in the context of classical (linear) elasticity theory
\cite{boussinesq, otto03}.  Thus the observation of response profiles
in two-dimensional disordered and hexagonal packings that differ
significantly on scales up to 30 grain diameters \cite{geng01, geng03}
requires consideration of nonlinear effects.  More generally, the
applicability of classical elasticity to granular media is a question
of ongoing research \cite{ellenbroek06, wyart05, blumenfeld02, gg05,
  tordesillas04, ostojic06}. 

Classical elasticity for an isotropic medium predicts a single-peaked
pressure profile that broadens linearly with depth \cite{boussinesq}.
Numerical results (see Ref.~\cite{gland05}, for example) demonstrate
responses well described by this solution in regions far from a
localized force in the bulk of a disordered frictional packing with
more than the critical number of contacts required for rigidity (the 
isostatic point). Recent work by Wyart \cite{wyart05} and Ellenbroek 
\cite{ellenbroek06} clarifies the onset of elastic behavior as average 
coordination number is increased above the isostatic limit.  
For materials with sufficiently strong uniaxial anisotropy, classical
elasticity theory admits double-peaked profiles with both peak widths
and the separation between peaks growing linearly as a function of
depth \cite{otto03}. The domain of applicability of classical
elasticity theory to granular materials is not well understood,
however, as it offers no simple way to incorporate noncohesive forces
between material elements or history dependent frictional forces.
Several alternative theories for granular stress response have been
proposed that make predictions qualitatively different from
conventional expectations.  Models of isostatic materials
\cite{tkachenko99, blumenfeld04} and models employing ``stress-only''
consititutive relations \cite{bouchaud97}, give rise to hyperbolic
differential equations for the stress and predict stress propagation
along characteristic rays.  Similarly, the directed force chain
network model predicts two diffusively broadening peaks developing
from a single peak at shallow depth \cite{socolar02}.  Numerical
studies in small isostatic or nearly isostatic packings also find
evidence of propagating peaks \cite{head01, kasahara04}.  Simulations
of weakly disordered hexagonal ball-and-spring networks, a common
example of an elastic material, can display two-peaked stress response
when the springs are one-sided \cite{gg02, gg05} and uniaxial
anisotropy is induced by contact breaking.  Response in the
ball-and-spring networks becomes single-peaked as friction increases,
a result mirrored by a statistical approach to hexagonal packings of
rigid disks \cite{ostojic05, ostojic06}.  Finally, a continuum
elasticity theory with a nonanalytic stress-strain relation at zero
strain has been shown to account quantitatively for single-peaked
stress response in rain-like preparations of granular layers \cite{liu06}.

We show here that an elasticity theory incorporating both hexagonal
anisotropy and near-field microstructure effects can account for the
experimental observations of Geng et al. \cite{geng01, geng03}
The theory is phenomenological; it accounts for the {\it average} 
stresses observed through a compilation of many individual response 
patterns. Our goal is to determine whether the ensemble average of 
effects of nonlinearities 
associated with force chains, contact breaking, and intergrain contact 
forces can be captured in a classical model, and, in particular, to 
account for the dramatic effects observed in experiments on 2D 
hexagonally close-packed systems. To that end, we develop a nonlinear 
continuum elasticity theory applicable to systems with hexagonal 
anisotropy \cite{ogden}. We find that these effects can account for the
quantitative discrepancy between the Boussinesq solution in 2D (the
Flamant solution) for linear systems and the experimental data of
Refs.~\cite{geng01} and \cite{geng03} for disordered packings of
pentagonal grains and hexagonal packings of monodisperse disks.  To
compare computed stress fields to the experimental data, we calculate
the pressure in the material as a function of horizontal position at
fixed depth.  We call such a curve a ``response profile.''

We find that induced dipole and quadrupole terms, which we attribute
to microstructure effects near the applied force, can account for the
narrowness of the response profiles in isotropic materials without
resorting to nonlinear effects.  In contrast, the response profiles
observed in hexagonal packings cannot be fit by the linear theory;
inclusion of nonlinear terms capable of describing hexagonal
anisotropy is required.  Using a theory based loosely on a simple
triangular lattice of point masses connected by springs, but allowing
an adjustable parameter specifying the degree of hexagonal anisotropy,
we find reasonable fits to the response profile data. We find that 
for sufficiently strong anisotropy the fitted response profiles 
correspond to small strains. Thus the nonlinear terms are necessary 
to capture the effects of material order, rather than large displacements. 
This is consistent with the experimental observations of Ref.~\cite{geng01}, 
for which the deformations were small and reversible.

The paper is organized as follows.  In Section~\ref{sec:review}, we
review well known elements of the theory of nonlinear elasticity and
the multipole expansion of the stress field.  In
Section~\ref{sec:energy}, we develop expressions for the free energies
of isotropic and several model hexagonal materials, including a model
in which strong nonlinearities arise for small strains. 
(We use the term ``free energy'' to maintain generality, though in the context of
granular materials, finite temperature effects are negligible and our
explicit models make no attempt to include entropic contributions.)
In Section~\ref{sec:nonlinear}, we present a perturbative expansion of
the response profiles for nonlinear systems in powers of inverse
distance from the point of application of the boundary force.  In
Section~\ref{sec:results}, we present the response profiles obtained
by adjusting the monopole, dipole, and quadrupole strengths and the
degree of hexagonal anisotropy.

\section{Review of elasticity concepts, definitions, and notation}
\label{sec:review}
\begin{figure}[tbp]
\centering
\includegraphics[clip,width=0.65\linewidth]{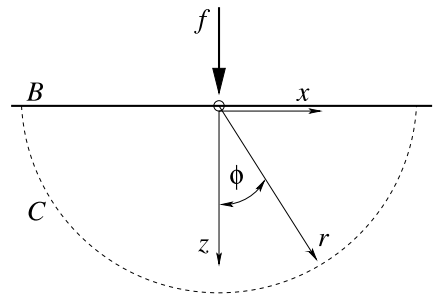}\\
\caption{Stress response in an elastic half-space. Forces must vanish everywhere on the 
free boundary, $B$, except at the origin. The total force transmitted across the surface 
$C$ is $f\hat{z}$. }
\label{fig:boussinesqproblem}
\end{figure}
We first provide a brief review of stress response in linear
elasticity theory for an isotropic half-plane.  We then describe the
general equations of nonlinear elasticity that are solved in Section~\ref{sec:nonlinear}
for particular forms of the free energy.  Finally, we review the multipole formalism
that is later used to model the effects of microstructure in the region near
the applied force where the continuum theory must break down.

The response of an elastic half-space to a point force normal to the
boundary, depicted in Fig.~\ref{fig:boussinesqproblem}, was first
given by Boussinesq \cite{boussinesq}. A normal force $f$ is applied
at the origin. In linear elasticity the stress components
$\sigma_{r\phi}$ and $\sigma_{\phi\phi}$ vanish on the surface $B$.
The force transmitted across a surface $C$ enclosing the boundary
force and with outward normal $\hat{n}$ must be equal to the force
applied at the boundary, namely $\int_C dC \, \hat{z} \cdot \cauchy
\cdot \hat{n} = f$. and $\int_C dC \, \hat{x} \cdot \cauchy \cdot
\hat{n} = 0$.  We expect that the Boussinesq result applies far from
the point of forcing, where the stress is weak and can be averaged
over a large representative volume of grains.  In this regime, the stress tensor
$\cauchy$ is solely radially compressive, independent of bulk and
shear moduli, and (in two dimensions) inversely proportional to the
distance from the point of application
\begin{equation}
\sigma_{rr} = \frac{2f\cos{\phi}}{\pi r}, \quad
\sigma_{r\phi} = 0, \quad 
\sigma_{\phi\phi}=0.
\label{eqn:bouss}
\end{equation}
Here $r$ and $\phi$ are polar coordinates, $\phi$ being measured from
the vertical as depicted in Fig.~\ref{fig:boussinesqproblem}.
Compressive stress is positive.  The stress contours are circles
passing through the origin, where the boundary force is applied. This
result is a useful approximation to the response in a real material
far from other boundaries.  For linear systems, it can be used to
calculate the response to an arbitrary distribution of force on the
boundary.

Nonlinearities arise from the proper geometric treatment of finite
strains and rotations as well as possible anharmonicity in the free
energy of the system. In classical elasticity, a linear constitutive
relation (e.g.~Hooke's law \cite{landau}) between stress and strain
results from a free energy $A$ that is quadratic in the components of
the strain tensor.  This can be regarded as the first term in a Taylor
expansion of $A$ about an equilibrium reference configuration, and in
this paper we include cubic and quartic contributions to the free
energy as well. Unlike the quadratic terms, the higher order
contributions can distinguish between a hexagonally anisotropic system
and an isotropic one.

When cubic and higher order powers of the strain in $A$ become
important, it may also be necessary to take into account geometric
sources of nonlinearity.  Let $\X = (X,Z)$ be the position of a
material element in the reference (undeformed) configuration and let
$\x = (x,z)$ be the position of the same material element in the
deformed configuration.  The displacement field is defined as $\U =
\x-\X$ and the deformation gradient is defined as
\begin{equation}
\F = {\bm 1}+\Grad\U,
\label{eqn:defgrad}
\end{equation}
where $\Grad\! = (\partial_X, \partial_Z)$.  To ensure invariance under
overall rotations, one must work with the full Lagrangian strain
\begin{equation} \label{eqn:lagrangianstrain}
\lagrangian = \frac{1}{2}\left(\F^\T \F - {\bm 1}\right)
\end{equation}
rather than just the
linearized strain $\linstrain = (\F^\T + \F)/2$.  In conventional
(linear) elasticity theory, the terms in $\lagrangian$ nonlinear in $\U$ are
neglected and $\Grad$ can be replaced by $\grad = (\partial_x, \partial_z)$.

The Cauchy stress $\cauchy$ is the stress measured in experiments and is a natural function of $\x$.  
It must satisfy the equations of force balance, 
$\divv \cauchy + \rho {\bm g} = 0$, and torque balance, $\cauchy^\T = \cauchy$, 
for any deformation. Here $\divv\!$ ($\Div\!$) is the divergence with 
respect to the deformed (undeformed) coordinates.  In the context of nonlinear models with boundary conditions 
expressed in terms of forces, these equations are more 
conveniently expressed with respect to the undeformed coordinates, the nominal stress
$\nominal = J \F^{-1} \cauchy$, and the reference density $\rho_0(\X) = J \rho(\x)$, where
$J = \det \F$.  The equations of force and torque balance can be rewritten 
\begin{eqnarray}
\Div \nominal + \rho_0 {\bm g} &=& 0,  \label{eqn:forcebalance}\\
(\F \nominal)^\T &=& \F \nominal. \label{eqn:torquebalance}
\end{eqnarray}

Defining the thermodynamic tension $\thermo$ via $\nominal= \thermo \F^\T$, the equations are closed by a constitutive relation coupling $\thermo$ to the Lagrangian strain (and through it the deformation gradient), namely $\thermo = \frac{\partial A}{\partial \lagrangian}$. Combining these, the nominal stress can be written as
\begin{equation}
\nominal = \frac{\partial A}{\partial \lagrangian}\F^\T.
\label{eqn:nominalstress}
\end{equation}
Together, Eqns.~(\ref{eqn:defgrad}-\ref{eqn:nominalstress}) represent a set of equations specifying the displacements in the system, for a specific material specified by the free energy $A$, and subject to the boundary conditions that stresses vanish on the deformed surface (except at the singular point) and the total force transmitted through the material is $f\hat{z}$.

Studies of the nonlinear Boussinesq problem have focused primarily on
stability analysis \cite{simmonds94, coon04, lee04}. Here we emphasize
the form of the stress response profile and restrict our attention to
two-dimensional isotropic and hexagonally anisotropic systems.  As
will be described below, the stress response can be developed in an
expansion in inverse powers of the distance from the boundary force,
reminiscent of a multipole expansion of an electromagnetic field.

\begin{figure}[tbp]
\centering
\includegraphics[clip,width=0.97\linewidth]{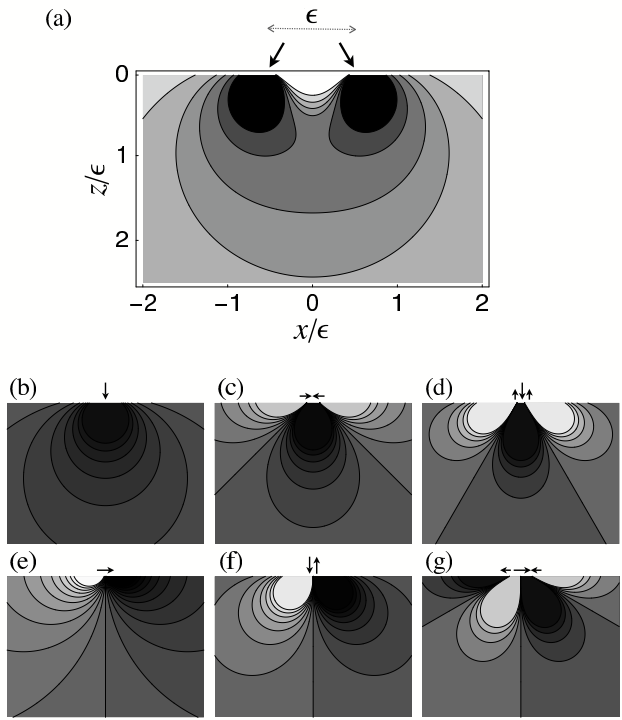}
\caption{(a) Contour plot of pressure for two point forces of equal magnitude located at 
$\pm \epsilon/2$ and oriented at $\pm \pi/6$ from the surface normal. Distances are in 
units of $\epsilon$. The response is two-peaked for shallow depths, transitioning to the 
circular contours of $\srr$ for a single normal force at the origin. Monopole (b,e), 
dipole (c,f), and quadrupole (d,g) boundary forcings, along with contours of the 
corresponding pressures. }
\label{fig:dipole_demo}
\end{figure}

The stress response of a hexagonal packing in Ref.~\cite{geng01}
(reproduced in Figs.~\ref{fig:springfit}-\ref{fig:betafit2}) displays
a rich structure, developing new peaks with increasing depth that
gradually broaden and fade.  It is apparent that Eq.~\eqref{eqn:bouss}
can never recreate such a response profile, as there is no length
scale over which the response develops.  However, it is possible to
create two- (or more) peaked response in isotropic linear elasticity.
All that is necessary is the application of more than one force at the
boundary.  Two boundary forces oriented at $\pm \pi/6$ to the normal,
for example, will produce a two-peaked stress response at shallow
depths, as shown in Fig.~\ref{fig:dipole_demo}a.  For depths much
greater than the distance between the two forces, the response
approaches that of a single normal force equal to the sum of the
normal components of the two boundary forces.

At distances larger than the separation between the points of
application of the force, the stress field in
Fig.~\ref{fig:dipole_demo}a can be closely approximated by a multipole
expansion.  In a granular material, the local arrangement of grains in
regions where strains are large will induce deviations from the
continuum theory, and in the Boussinesq geometry the far field effects
of these deviations can be approximated by placing a series of
multipolar forcing terms at the origin.  Thus, although the physical
force applied by Geng et al., for example, was a single, sharply
localized, normal force, 
we include in our continuum theory parameters specifying dipole, quadrupole, and perhaps higher order multipole forcing strengths to account for the effect of microstructure.
If the applied 
force is spread over enough grains that the continuum solution
predicts only small strains everywhere, then the multipole contributions
can be explicitly computed within the continuum theory.  If, on the other hand, the force is applied
to a single grain and represented as a delta-function in the continuum
theory, the theory will predict large strains near the origin and
microstructure effects must be taken into account either phenomenologically, as we do here, or through a more detailed model of the microstructure in the vicinity of the applied force.
We conjecture that the size of this region near the origin scales with the 
``isostaticity length scale'' discussed in Refs.~\cite{wyart05} and \cite{ellenbroek06}.

The first several multipole forces and corresponding pressure
profiles, are depicted in Fig.~\ref{fig:dipole_demo}b-g. A multipole
force with stresses that decay as $1/r^n$ can be constructed from
$n$ evenly spaced compressive or shearing boundary forces having
alternating directions and magnitudes in proportion to the
$n^\mathrm{th}$ row of Pascal's Triangle. The integral
$\int_{-\infty}^{\infty}dx\,x^{n-1} {\bm f}(x)$ is the lowest order
nonvanishing moment of the boundary force distribution ${\bm f}(x)$.

The form of the far-field stress response to multipole forcing in linear elasticity 
can be developed by considering the Airy stress function $\chi$ such that 
$\sigma_{rr}=\partial_r \chi/r + \partial_{\phi\phi} \chi/r^2$,   
$\sigma_{r\phi} = \sigma_{\phi r} = -\partial_r(\partial_\phi \chi/r)$, and
$\sigma_{\phi\phi} = \partial_{rr}\chi$. The Airy 
stress function is biharmonic:
\begin{equation}
\bigtriangleup\bigtriangleup\chi = 0.
\label{eqn:biharmonic}
\end{equation}
Assuming $\chi$ has the form
\begin{equation}
\chi(r,\phi) = r^2 \sum_{n=1}^\infty \frac{1}{r^n} \chi^{(n)}(\phi)
\end{equation}
and solving for $\chi^{(n)}$ yields a series of corresponding tensors $\cauchy^{(n)}$. 
(It is convenient to restrict ourselves to transversely symmetric multipole terms, 
such as those in Fig.~\ref{fig:dipole_demo}b-d, so that there is only one corresponding 
stress tensor for each value of $n$.) $\cauchy^{(1)}$ corresponds to the monopole of 
Eq.~(\ref{eqn:bouss}).
For each $\cauchy^{(n)}$, $\sigma^{(n)}_{\phi\phi}$ and $\sigma^{(n)}_{r\phi}$ must 
vanish on the surface except at the origin. 
For the surface $C$ in Fig.~\ref{fig:boussinesqproblem} we generalize the monopole normalization 
to arbitrary $n$:
\begin{eqnarray}
\int_{-\frac{\pi}{2}}^{-\frac{\pi}{2}}r d\phi\,(r \sin{\phi})^{n-1} 
\left( \hat{p}\cdot \cauchy^{(n)} \cdot \hat{r}\right) &=& 0 
\nonumber \\
\int_{-\frac{\pi}{2}}^{-\frac{\pi}{2}}r d\phi\,(r \sin{\phi})^{n-1}
\left( \hat{q}\cdot \cauchy^{(n)} \cdot \hat{r}\right) &=& ka^n,
\end{eqnarray}
where $\hat{p} = \hat{x}$ ($\hat{z}$) and $\hat{q} = \hat{z}$ ($\hat{x}$) for odd (even) 
powers of $n$.
$k$ and $a$ carry the units of stress and length, respectively, the
applied force is $f=ka$.
Subject to this normalization, the dipole stress tensor $\cauchy^{(2)}$ is
\begin{eqnarray}
\sigma^{(2)}_{rr} &=& \frac{8 ka^2}{\pi r^2}\cos{2\phi} \nonumber \\
\sigma^{(2)}_{r\phi} &=& \frac{4 ka^2}{\pi r^2}\sin{2\phi} \nonumber \\
\sigma^{(2)}_{\phi\phi} &=& 0,
\label{eqn:dipolestress}
\end{eqnarray}
and the quadrupole stress tensor $\cauchy^{(3)}$ is
\begin{eqnarray}
\sigma^{(3)}_{rr} &=& - \frac{5 ka^3}{\pi r^3}\cos{3\phi} -\frac{3 ka^3}{\pi r^3}\cos{\phi}  
\nonumber \\ 
\sigma^{(3)}_{r\phi} &=& - \frac{3 ka^3}{\pi r^3}\sin{3\phi} -\frac{3 ka^3}{\pi r^3}\sin{\phi} 
\nonumber \\
\sigma^{(3)}_{\phi\phi} &=& \hphantom{-}\frac{ ka^3}{\pi r^3}\cos{3\phi} + \frac{3 ka^3}{\pi r^3}\cos{\phi}.
\label{eqn:quadrupolestress}
\end{eqnarray}
Contours of the associated pressures $p^{(n)}=(1/2)\Tr\cauchy^{(n)}$ and sample boundary forces which 
produce them are shown in Fig.~\ref{fig:dipole_demo}b-d.

The higher order multipole terms decay more quickly than the 
monopole term, so at asymptotically large depth in a material in which both monopole and higher order terms 
are present, the response is indistinguishable from the 
Boussinesq solution.  Closer to the point of application, the induced 
multipole terms contribute more complex structure to the response. The distance over 
which this structure is observable depends on the material properties through the elastic coefficients 
and increases with the strength of the applied force $f$. 

\section{Model free energies}
\label{sec:energy}
Here we develop expressions for the elastic free energy of several model systems having hexagonal 
symmetry.  These will be needed to construct constitutive relations relating 
stress and strain. 

\subsection{Symmetry considerations}
To linear order the elastic energy is quadratic in the strain components:
\begin{equation}
A = \frac{1}{2}\lambda_{ijkl}\, \eta_{ij}\,\eta_{kl}.
\end{equation}
$\lambda$ is a fourth order tensor of rank two, and its components are the elastic coefficients 
of the material. For an isotropic 
material the free energy must be invariant for rotations of $\eta$ through arbitrary angle. 
Therefore $A$ can depend only on scalar functions of the strain tensor components. 
In two dimensions, the strain tensor has two eigenvalues or principal invariants. All other 
scalar invariants, including the independent invariants $I_1 = \Tr \lagrangian = \eta_{ii}$ 
and $I_2 = \Tr \lagrangian^2 = (\eta_{ij})^2$ (summation implied), can be expressed in 
terms of the principal invariants \cite{spencer} or, equivalently, in terms of $I_1$ and $I_2$.
The free energy of an isotropic linear elastic material can be expressed in terms of 
combinations of $I_1$ and $I_2$ that are quadratic in the strain components.
\begin{equation}
A = \frac{1}{2}\lambda I_1^2 + \mu I_2
\end{equation}
where $\lambda$ and $\mu$ are the Lam\'e coefficients. 
The reasoning generalizes to higher orders. At each order, 
there will be as many elastic coefficients as there are independent combinations of $I_1$ 
and $I_2$.  To quartic order in the strains, we have
\begin{eqnarray}
A &=& 
\hphantom{+\quad} \biggl(
\frac{1}{2}\lambda I_1^2 + \mu I_2 \biggr) \nonumber \\
&&    + \quad   \biggl( \omega_1 I_1^3 + \omega_2 I_1 I_2 \biggr)  \nonumber \\
&&    + \quad   \biggl(\Omega_1 I_1^4 + \Omega_2 I_2^2 + \Omega_3 I_1^2 I_2 \biggr).
\label{eqn:isotropicenergy}
\end{eqnarray}
We refer to the $\omega$'s and the $\Omega$'s as third and fourth order elastic 
coefficients, respectively.

To construct the free 
energy of a hexagonal material, it is useful to consider a change of coordinates
\begin{eqnarray}
\xi &=& x+iz \nonumber \\
\zeta &=& x-iz,
\end{eqnarray}
as suggested in Ref.~\cite{landau}.
For a rotation of $\pi/3$ about $(\hat{z}\times\hat{x})$ these coordinates transform as
$\xi \rightarrow \xi e^{\pi i/3}$ and $\zeta \rightarrow \zeta e^{-\pi i/3}$. The free 
energy of an elastic material must be invariant under such a rotation, which implies that a component of 
the tensor $\lambda$ can be nonzero if and only if it too is invariant. 
For example, the quadratic coefficient $\lambda_{\xi \xi \zeta \zeta}$ is nonzero 
because, under rotation by $\pi/3$, $\lambda_{\xi \xi \zeta \zeta}
\rightarrow e^{\pi i/3} e^{\pi i/3} e^{-\pi i/3} e^{-\pi i/3} 
\lambda_{\xi \xi \zeta \zeta} = \lambda_{\xi \xi \zeta \zeta}$. The only other 
independent nonzero quadratic coefficient is $\lambda_{\xi  \zeta \xi \zeta}$. 
Cubic and higher order coefficients, which are labeled by six or more indices, can also be invariant 
by having six like indices, as in $\lambda_{\xi \xi \xi \xi \xi \xi}$.  
There are three independent coefficients at cubic order and four at quartic order. 

The general form of the free energy of a hexagonal material is, to quartic order, 
\begin{eqnarray}
A &=& 
\hphantom{+\quad}
\frac{1}{2!}\biggl(2 \lambda_1 \eta_{\xi \xi} \eta_{\zeta \zeta} + 
4 \lambda_2 \eta_{\xi \zeta}^2 \biggr)  \nonumber \\
&& + \quad \frac{1}{3!}\biggl(\Lambda_1 (\eta_{\xi\xi}^3+\eta_{\zeta \zeta}^3) + 
12 \Lambda_2 \eta_{\xi\xi}\eta_{\xi \zeta}\eta_{\zeta \zeta} + 
8 \Lambda_3 \eta_{\xi \zeta}^3\biggr)  \nonumber \\
&& + \quad \frac{1}{4!}\biggl( 
6 L_1\eta_{\xi\xi}^2\eta_{\zeta \zeta}^2 +
48 L_2 \eta_{\xi\xi}\eta_{\zeta \zeta}\eta_{\xi \zeta}^2 \biggr. \nonumber \\ 
&& \quad \quad \quad \quad \biggl.
+16 L_3 \eta_{\xi \zeta}^4 + 
8 L_4(\eta_{\xi\xi}^3\eta_{\xi \zeta} + 
\eta_{\zeta \zeta}^3\eta_{\xi \zeta}) 
\biggr)
\label{eqn:hexagonalenergy}
\end{eqnarray}
where 
$\eta_{\xi \xi} = \eta_{xx} - \eta_{zz} + 2i\eta_{xz}$, 
$\eta_{\zeta \zeta} = \eta_{xx} - \eta_{zz} - 2i\eta_{xz}$, and
$\eta_{\xi \zeta} = \eta_{xx}+\eta_{zz}$.

For simplicity, we have assumed that terms involving gradients of the strains
are negligible \cite{suiker01, walsh04}. 

\subsection{Hexagonal ball-and-spring network}
\begin{figure}[tbp]
\centering
\includegraphics[clip,width=0.9\linewidth]{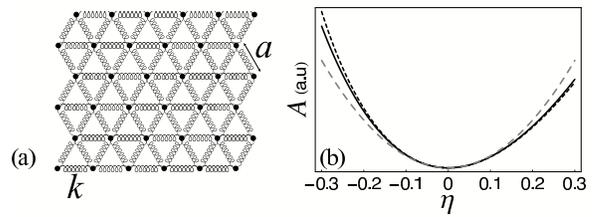}
\caption{(a) A ball-and-spring network with hexagonal symmetry and springs oriented 
  horizontally. Even for a linear force law, the free energy has terms
  of cubic and higher order in the strains when the equilibrium length
  of the springs is nonzero. (b) Free energy as a function of strain
  for a unit cell of the ball-and-spring network in (a).  (solid
  black) Vertical uniaxial compression: $\eta = \eta_{zz}$ with
  $\eta_{yy} = 0$. (dashed black) $\eta = \eta_{xx}$ with Horizontal
  uniaxial compression: $\eta_{zz} = 0$. (dashed gray) Linear elastic
  approximation for both cases. $\eta<0$ corresponds to compression. }
\label{fig:ballspring}
\end{figure}

We now construct the free energy for several specific hexagonal materials, taking the 
point-mass-and-spring network of Fig.~\ref{fig:ballspring}a as a starting point.  The elastic 
coefficients are determined by calculating the free energy under a homogeneous strain and comparing to 
Eq.~\eqref{eqn:hexagonalenergy}. 
The springs are taken to have an equilibrium length $\ell$ and to obey Hooke's law: 
for a spring with one end at $\x_1$ and the other at $\x_2$ the force is
\begin{equation}
f=-k\left(\sqrt{(\x_2-\x_1)\cdot(\x_2-\x_1)} - \ell\right),
\end{equation}
where $k$ is the spring constant.  
We take the springs to be at their equilibrium lengths in the
undeformed system: $\ell=a$, the lattice constant.  

Consider the homogeneous strain
\begin{equation}
\eta = \left(
  \begin{array}{cc}
  \eta_{xx} & 0 \\
  0 & \eta_{zz} 
  \end{array}\right)
\end{equation}
which stretches the coordinates to $x = \sqrt{1+2\eta_{xx}}\,X$ and 
$z = \sqrt{1+2\eta_{zz}}\,Z$. The free energy per unit (undeformed) 
volume of a hexagonal ball-and-spring network with one third of the springs oriented along the $\hat{x}$ 
direction under this stretch is
\begin{eqnarray}
\frac{4}{\sqrt{3} k}A & =& 
\hphantom{+\quad} \biggl(
\frac{3}{2}\eta_{xx}^2 + \frac{3}{2}\eta_{zz}^2 
+ \eta_{xx}\eta_{zz} 
\biggr)  \nonumber \\
&& - \quad \biggl(
+ \frac{9}{8}\eta_{xx}^3 + \frac{11}{8}\eta_{zz}^3
+\frac{9}{8}\eta_{xx}^2\eta_{zz} + \frac{3}{8}\eta_{xx}\eta_{zz}^2 
\biggr) \nonumber \\
&& + \quad \biggl(
\frac{135}{128}\eta_{xx}^4 
+ \frac{215}{128}\eta_{zz}^4  
+\frac{45}{64}\eta_{xx}^2\eta_{zz}^2 
\biggr. \nonumber \\
&& \hphantom{\biggl( \quad \quad \quad \quad \quad} \biggl.
+\frac{45}{32}\eta_{xx}^3\eta_{zz} 
+ \frac{5}{32}\eta_{xx}\eta_{zz}^3 
\biggr).
\label{eqn:ballspringenergy}
\end{eqnarray}
The presence of cubic and higher order terms in 
the free energy is due to the nonzero spring equilibrium length.
The free energy for a constrained axial compression/extension 
in the $\hat{x}$ and $\hat{z}$ directions is plotted in Fig.~\ref{fig:ballspring}. The 
corrections to the quadratic expression stiffen the system under compression and soften it 
slightly under small extensions.

Comparing Eqs.~\eqref{eqn:hexagonalenergy} and \eqref{eqn:ballspringenergy} and 
equating like coefficients of $\eta_{xx}$ and $\eta_{zz}$ we find
\begin{equation}
\begin{array}{ll}
\vspace{0.1in}
\lambda_1 = \lambda_2 = \frac{\sqrt{3}}{8}k  \\
\vspace{0.1in}
\Lambda_1 = -\Lambda_2 = -\Lambda_3 = \frac{3\sqrt{3}}{32}k  \\
 L_1 = L_2 = L_3 = -L_4 = \frac{15\sqrt{3}}{128}k
 \label{eqn:springcoeffs}
\end{array}
\end{equation}
A similar calculation for a material in which one third of the springs are oriented vertically,
corresponding to a reference configuration rotated by $90^{\circ}$ from the one shown in
Fig.~\ref{fig:ballspring}, yields
\begin{equation}
\begin{array}{ll}
\vspace{0.1in}
\lambda_1 = \lambda_2 = \frac{\sqrt{3}}{8}k  \\
\vspace{0.1in}
-\Lambda_1 = -\Lambda_2 = -\Lambda_3 = \frac{3\sqrt{3}}{32}k  \\
L_1 = L_2 = L_3 = L_4 = \frac{15\sqrt{3}}{128}k 
\end{array}
\end{equation}

\subsubsection{The $\alpha$-material}
\begin{figure}[tbp]
\centering
\includegraphics[clip,width=0.65\linewidth]{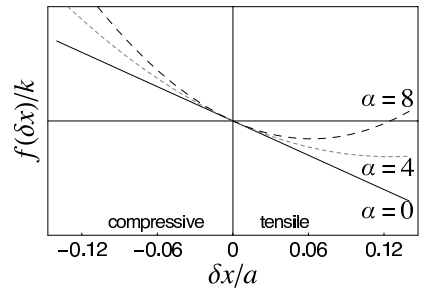}
\caption{The force law of Eq.~\eqref{eqn:tensilesoftening} for $k=1$ and 
$\alpha = 0\ldots 8$. }
\label{fig:alphaforcelaw}
\end{figure}
Goldenberg and Goldhirsch \cite{gg02, gg04, gg05} find two-peaked stress 
response in numerical simulations of a hexagonal lattice of springs when 
the springs are allowed to break under tensile loading.  
Contact-breaking explicitly breaks our assumption of local hexagonal anisotropy 
in any particular sample. In the context of an ensemble average, however,
the material description retains hexagonal symmetry and the effects of
contact breaking are captured phenomenologically by considering
material made of springs with a force law that softens under extension.

\begin{eqnarray}
f &=&
-k\left(\sqrt{(\x_2-\x_1)\cdot(\x_2-\x_1)} - a\right) \nonumber \\
&& \quad \quad
+ k\frac{\alpha}{a} \left(\sqrt{(\x_2-\x_1)\cdot(\x_2-\x_1)} - a\right)^2.
\label{eqn:tensilesoftening}
\end{eqnarray}
For $\alpha>0$ the springs soften under tension and stiffen under compression, as shown in 
Fig.~\ref{fig:alphaforcelaw}. In the
horizontal orientation the elastic constants from Eq.~\eqref{eqn:springcoeffs} are 
shifted according to 
\begin{equation}
\begin{array}{ll}
\vspace{0.1in}
\lambda_1 = \lambda_2 = \frac{\sqrt{3}}{8}k  \\
\vspace{0.1in}
\Lambda_1 = -\Lambda_2 = -\Lambda_3 = \frac{3\sqrt{3}}{32}k - \frac{\alpha}{a}\frac{3\sqrt{3}}{32}k \\
L_1 = L_2 = L_3 = -L_4 = \frac{15\sqrt{3}}{128}k - \frac{\alpha}{a}\frac{9\sqrt{3}}{64}k
\end{array}
\end{equation}

\subsubsection{The $\beta$-material}
In the spirit of phenomenological modeling, all of the elastic constants consistent with
hexagonal symmetry should be considered to be parameters to be determined by experiment.
To probe the importance of hexagonal anisotropy, we consider a model in which all elastic
constants but one are fixed and define a parameter $\beta$ corresponding to the strength
of the anisotropy.
Note that the elastic constants for the two orientations of the hexagonal 
ball-and-spring network considered above can be rewritten as
\begin{equation}
\begin{array}{ll}
\vspace{0.1in}
\lambda_1 = \lambda_2 = \frac{\sqrt{3}}{8}k  & \\
\vspace{0.1in}
\Lambda_2 = \Lambda_3 = -\frac{3\sqrt{3}}{32}k &
\quad \Lambda_1 = \beta \frac{3\sqrt{3}}{32}k \\
L_1 = L_2 = L_3 = \frac{15\sqrt{3}}{128}k &
\quad L_4=  -\beta \frac{15\sqrt{3}}{128}k .
\end{array}
\label{eqn:betamaterial}
\end{equation}
The case $\beta=1$ gives the network with horizontal springs; $\beta=-1$ gives the 
network with vertical springs; and $\beta = 0$ gives an isotropic system. Linear response for 
elastic materials with other ansisotropies is treated in Ref.~\cite{otto03}.

\section{Method} \label{sec:nonlinear}
We wish to obtain corrections to the linear elastic result 
for a material with hexagonal symmetry. For later convenience we write $f=\Qm ka$, where 
$\Qm$ is dimensionless, $k$ has units of a spring constant, and $a$ is a lattice constant with 
units of length. We expand the stress in successive inverse powers of the radial coordinate, 
and refer to the terms in the expansion as the dipole correction, quadrupole correction, and 
so forth.  For simplicity and clarity, we present here the calculation corresponding to
the free energy of Eq.~\eqref{eqn:hexagonalenergy} with coefficients given in
Eq.~(\ref{eqn:springcoeffs}) in detail.  General equations for arbitrary elastic coefficients
are exceedingly long and unilluminating.  
 
We solve for the the displacements 
 $u_R(R,\Phi)$ and $u_\Phi(R,\Phi)$, from which the stress tensor can be reconstructed.
Capitalized coordinates are used as we are now 
 careful to distinguish between the deformed and undeformed states. After the deformation, 
 the point $\X$ is at 
 $\x = \X + u_R(R,\Phi))\hat{R} + u_\Phi(R,\Phi)\hat{\Phi}$. 
To linear order and for 
 the ball-and-spring network described in Eq.~\eqref{eqn:springcoeffs} the displacements are
 \begin{eqnarray}
 u_R^{(0)}(R,\Phi) &=& 
 \frac{\sqrt{3}\Qm a}{\pi} \left( 
 \cos{\Phi}\lnR  + \frac{1}{3}\Phi \sin{\Phi} \right); \nonumber \\
 u_\Phi^{(0)}(R,\Phi) &=& 
 \frac{\sqrt{3}\Qm a}{\pi} \left(\vphantom{\frac{1}{3}}
 -\sin{\Phi}\lnR \right.\nonumber \\
&& \hphantom{\frac{\sqrt{3}\Qm a}{\pi} \biggl( \, } \left.
 -\frac{2}{3} \sin{\Phi} + \frac{1}{3} \Phi \cos{\Phi}\right).
\label{eqn:lineardisplacements}
 \end{eqnarray}

The parameter $R_0$ requires comment. Because the material is semi-infinite 
in extent, it is free to undergo an arbitrary rigid-body translation in the 
$\hat{Z}$-direction under the influence of a normal boundary force. 
Thus the point along the $Z$-axis at which the deformation $\U$ is 
zero may be chosen arbitrarily. $R_0$ parameterizes this 
variation. 
Note that the nominal stress, which in the linear theory is equivalent to 
$\cauchy$ in Eq.~(\ref{eqn:bouss}), is independent of $R_0$.

To find the dipole correction, we take $u_R = u_R^{(0)}+u_R^{(1)}$ and 
$u_\Phi = u_\Phi^{(0)}+u_\Phi^{(1)}$ and assume a correction of the form
\begin{eqnarray}
u_R^{(1)}(R,\Phi) &=& a^2\frac{v_0(\Phi)}{R} + 
a^2\frac{v_1(\Phi)}{R}\lnR \nonumber \\
u_\Phi^{(1)}(R,\Phi) &=& a^2\frac{w_0(\Phi)}{R} + 
a^2\frac{w_1(\Phi)}{R}\lnR.
\end{eqnarray}
The deformation gradient $\F$ in polar coordinates is 
\begin{equation}
 \F = \left(
   \begin{array}{cc}
   1+\partial_R u_R & \left(\partial_\Phi u_R - u_\Phi\right)/R \\
   \partial_R u_\Phi & 1+\left(\partial_\Phi u_\Phi + u_R \right)/R
   \end{array}\right).
\label{eqn:polarF}
\end{equation}
Through Eqs.~\ref{eqn:lagrangianstrain} and \ref{eqn:nominalstress} 
the nominal stress can be written entirely in terms of the displacements, and through them 
in terms of the four unknown functions, $v_0$, $v_1$, $w_0$, and $w_1$. 

Substituting the linear Boussinesq solution of Eq.~\ref{eqn:lineardisplacements} in
Eq.~\ref{eqn:polarF}, evaluating Eq.~(\ref{eqn:forcebalance}), and requiring the coefficient
of $1/R^3$ to vanish yields conditions on the $v$'s and $w$'s. 
(Terms of smaller order in $1/R$ vanish identically.)
We find
\begin{eqnarray}
&&\!\!\!\!\!\!
11 - 13\cos{2\Phi} - 3\cos{4\Phi} - 9\cos{6\Phi} - 6\cos{8\Phi} \nonumber \\
& &\quad = 
\frac{9}{2}v_0^{\prime\prime}-27v_1 -27w_0^\prime +9w_1^\prime \nonumber \\
& &\quad \hphantom{=}
+\left(\frac{9}{2}v_1^{\prime\prime}-27w_1^\prime\right)\lnR; \nonumber \\
&&\nonumber \\
&&\!\!\!\!\!\!
 -5\sin{2\Phi} + \sin{4\Phi} + 3\sin{6\Phi} + 2\sin{8\Phi} \nonumber \\
& &\quad = 
3v_0^\prime + 3v_1^\prime + \frac{9}{2}w_0^{\prime\prime}-3w_1 \nonumber \\
& &\quad \hphantom{=}
+ \left(3v_1^\prime + \frac{9}{2}w_1^{\prime\prime}\right)\lnR.
\label{eqn:firstode}
\end{eqnarray}
For the moment, we neglect terms of higher order in $1/R$. The source terms on the 
left-hand side in Eq.~\eqref{eqn:firstode} are generated by the linear solution. 
Requiring coefficients of $\ln R$ to vanish independently gives four 
second-order ordinary differential equations for the four unknown functions. 

The conditions that normal and shear forces vanish everywhere on the deformed boundary except at the point of application of the external force can be written
\begin{eqnarray}
S_{\Phi R}(R \neq 0, \Phi = \pm \pi/2) &=& 0 \nonumber \\
S_{\Phi\Phi}(R \neq 0, \Phi = \pm \pi/2) &=& 0.
\label{eqn:boundaryconditions}
\end{eqnarray}
Both the $S_{\Phi R}$ and $S_{\Phi\Phi}$ components of stress have terms proportional to $\ln{R}$. When we require these terms to vanish independently of all other terms, Eq.~(\ref{eqn:boundaryconditions}) represents eight constraints.
The nominal stress must also satisfy force-transmission conditions
\begin{eqnarray}
\int_{C}\,\hat{x}\cdot\nominal^T \cdot\hat{n}\,dC &=& 0 \nonumber \\
\int_{C}\,\hat{z}\cdot\nominal^T \cdot\hat{n}\,dC &=& f,
\label{eqn:secondbc}
\end{eqnarray}
where $C$ is any surface enclosing the origin (see e.g.~Fig.~\ref{fig:boussinesqproblem}) 
and $\hat{n}$ is the unit normal to $C$. Eq.~\eqref{eqn:secondbc} is satisfied 
by the linear elastic solution, and all solutions to Eq.~\eqref{eqn:firstode} subject to 
Eq.~(\ref{eqn:boundaryconditions}) contribute zero under the integration, 
so this provides no additional constraint on the system. 

The eight constraints of Eq.~(\ref{eqn:boundaryconditions}) fix only seven 
of the eight integration constants. The eighth integration constant, which 
we label $\Qd$, multiplies terms identical 
to those contributed in linear elasticity by a horizontally oriented dipole 
forcing such as that depicted in Fig.~\ref{fig:dipole_demo}c and given in 
Eq.~(\ref{eqn:dipolestress}). $\Qd$ is fixed by 
demanding that a variation of the parameter $R_0$ produce only a rigid body 
translation of the material. 
The integration constants determined in this way  produce a nominal stress 
$\nominal$ independent of $R_0$, as must be the case.

\begin{figure}[tbp]
\centering
\includegraphics[clip,width=0.65\linewidth]{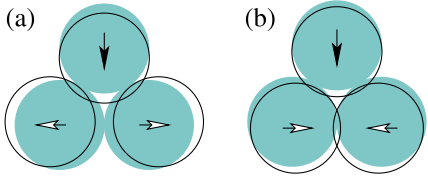}
\caption{Two imagined scenarios in which a point force induces a dipole. Regions of 
overlap indicate a compressive contact. (a) The 
disks in the second layer slide outward, e.g.~for low friction. (b) Alternatively the disks 
might roll inward, towards the line of force,  e.g.~due to greater friction between grains. 
This would select a dipole term in the stress response with opposite sign from the case 
depicted in (a). Thus, the details of the near field response depend on the mechanics of the 
discrete system. }
\label{fig:induceddipole}
\end{figure}

The solution of 
Eq.~\ref{eqn:firstode} consistent with Eq.~\eqref{eqn:boundaryconditions} is
\begin{eqnarray}
v_0(\Phi) &=&
 \left(\frac{\Qm^2}{\pi^2 } \right)
 \left[
 \frac{5}{6}  + \frac{7}{3}\cos{2\Phi}
 \right. \nonumber \\
 & & \,\, \hphantom{\left(\frac{\Qm^2 }{\pi^2 } \right)}
 + \frac{1}{4}\cos{4\Phi} + \frac{1}{4}\cos{6\Phi}  \nonumber \\
 & & \,\,  \hphantom{\left(\frac{\Qm^2 }{\pi^2 } \right)} \left. 
 + \frac{1}{12}\cos{8\Phi}+\frac{11}{6}\Phi \sin{2\Phi}
  \right]   \nonumber \\
 & & -\left( \frac{4\pi\Qd}{\sqrt{3}} \right)\cos{2\Phi}; \nonumber \\
% \end{eqnarray}
% \begin{eqnarray}
v_1(\Phi) &=& \left(\frac{3\Qm^2 }{2\pi^2 } \right)
 \cos{2\Phi} \lnR; \nonumber \\
% \end{eqnarray}
% \begin{eqnarray}
w_0(\Phi) &=&
-\left(\frac{\Qm^2 }{\pi^2 } \right) \left[
\frac{11}{9}\Phi  + \frac{2}{3}\sin{2\Phi} \right. \nonumber \\
 & &  \,\, \hphantom{-\left(\frac{\Qm^2 }{\pi^2 } \right)}
 +\frac{1}{12}\sin{4\Phi}
 +\frac{1}{12}\sin{6\Phi}   \nonumber \\
 & & \,\, \hphantom{-\left(\frac{\Qm^2 }{\pi^2 } \right)}
 \left.
 + \frac{1}{36}\sin{8\Phi} - \frac{11}{18}\Phi\cos{2\Phi} 
 \right] \nonumber \\
& & + \left( \frac{4\pi\Qd}{3\sqrt{3}} \right) \sin{2\Phi} \nonumber \\
% \end{eqnarray}
% \begin{eqnarray}
w_1(\Phi) &=& -\left(\frac{\Qm^2}{2\pi^2}\right)\sin{2\Phi}\lnR.
\end{eqnarray}

For the choice $R_0=a$, we find the induced dipole coefficient $\Qd=0$, 
and for the sequel we fix $R_0$ to have this value. The same choice of $R_0$
also yields the induced quadrupole coefficient $\Qq=0$ below. As discussed 
above, rather than set them to zero, we leave these terms in the 
displacements, and correspondingly the stresses, as free parameters to 
account for the influence of microstructure on the response. They are 
weighted so that $\Qd=1$ and $\Qq=1$  correspond to the stresses of 
Eqs.~(\ref{eqn:dipolestress}) and (\ref{eqn:quadrupolestress}).

We repeat the process described above to develop quadrupole  
corrections to the stress response. The displacements are assumed to have the form
$u_R(R,\Phi) = u_R^{(0)}(R,\Phi) + u_R^{(1)}(R,\Phi) +  u_R^{(2)}(R,\Phi)$ 
and 
$u_\Phi(R,\Phi) = u_ \Phi ^{(0)}(R,\Phi) + u_ \Phi ^{(1)}(R,\Phi) +  u_ \Phi ^{(2)}(R,\Phi)$
where the second order corrections have the form
\begin{eqnarray}
 \nonumber \\
 u_R^{(2)}(R,\Phi) &=& 
 \frac{V_0(\Phi)}{R^2} + 
\frac{V_1(\Phi)}{R^2}\lnR \nonumber \\
&& +\frac{V_2(\Phi)}{R^2}\lnR^2; \nonumber \\
u_ \Phi^{(2)}(R,\Phi)  &=&
\frac{W_0(\Phi)}{R^2} + 
\frac{W_1(\Phi)}{R^2}\lnR \nonumber \\
&& +\frac{W_2(\Phi)}{R^2}\lnR^2.
\end{eqnarray}
The details of the calculation are omitted, as they are conceptually similar to the dipole
calculation but involve much longer expressions. Defining $c_n \equiv \cos{n\Phi}$, 
$s_n \equiv \sin{n\Phi}$, and $L \equiv \ln{(R/R_0)}$, the pressure is
\begin{widetext}
\begin{eqnarray}
p(r(R,\Phi),\phi(R,\Phi)) 
&=& 
\frac{\Qm k a}{\pi} \frac{c_1}{R} 
- \frac{4\Qd k a^2}{ \pi} \frac{c_2}{R^2} 
- \frac{2\Qq k a^3}{\pi} \frac{c_3}{R^3} \nonumber \\
&& + \frac{B_2}{R^2}\biggl[17 -c_2 - 6c_4 - 9c_6 - 4c_8 - 22\Phi s_2 - 18 L c_2  \biggr] \nonumber \\
&& + \frac{B_3}{R^3}\biggl[
-\frac{99}{2}c_1 + \left(\frac{616277}{8820} - \frac{27}{7}\pi^2 + \frac{41}{3}\Phi^2 \right)c_3
+\frac{205}{2}c_5 + \frac{139}{3}c_7 + 25c_9  \biggr. \nonumber \\
&& \hphantom{\frac{B_3}{R^3}\biggl[}
+ \frac{63}{4}c_{11} + \frac{119}{10}c_{13} + \frac{10}{3}c_{15}  
- \Phi\left( 66s_1 + 161 s_3 - 66 s_5 - 88 s_7 - \frac{110}{3} s_9 \right) \biggr. \nonumber \\
&& \hphantom{\frac{B_3}{R^3}\biggl[}
+ L \left( -48c_1 - \frac{329}{3}c_3 + 36 c_7 + 30 c_9 - 42 \Phi s_3 \right) 
- 27L^2 c_3 \biggr] \nonumber \\
&& +\frac{B_3^{\prime}}{R^3} \Qd 
\biggl[ 8c_1 - \frac{151}{14}c_3 - 6c_7 - 5c_9 + 7\Phi s_3 + 9L c_3 \biggr],
\label{eqn:pressure}
\end{eqnarray}
\end{widetext}
where $B_2 = \Qm^2 k a^2/12 \sqrt{3}/\pi^2$, $B_3 = \Qm^3 k a^3/36\pi^3$, and 
$B_3^\prime=4\Qm^3 k a^3/3\sqrt{3}\pi^2$.

We will find below that the $\beta$-material best describes the data 
of Ref.~\cite{geng01}. In this case the pressure of 
Eq.~(\ref{eqn:pressure}) gains a number of additional terms involving 
$\beta$. These terms are given in the Appendix.

\section{Results} \label{sec:results}
Given the expressions derived above for the pressure, we perform numerical fits to the data from Geng et al.~\cite{geng01}.
There are four fitting parameters for the ball-and-spring material: the monopole coefficient 
$\Qm$, the dipole coefficient $\Qd$, the quadrupole coefficient $\Qq$, and the spring constant 
$k$. We take the lattice constant to be the disk diameter: $a=0.8\,$cm. The three multipole 
coefficients have been defined to be dimensionless.  
We set $R_0=a$ so that $\Qd$ and $\Qq$ would be zero in a theory with no 
microstruture correction.
In two dimensions the units of stress are the same as the units of the 
spring constant $k$. Thus $k$ sets the overall scale for the stress. 
For theoretical purposes, $k$ could be scaled to unity; in our fits it serves
merely to match the units of stress in the experimental data.

We attempt to fit experimental measurements on pentagonal-grain packings by 
varying $Q_m$, $Q_d$ and $Q_q$ in the isotropic theory. To explain the 
experimental data on hexagonal disk packings, we attempt fits based on 
the ball-and-spring network, the $\alpha$-material, and the $\beta$-material.

We regard the response profiles presented in the following section, particularly 
Figs.~\ref{fig:isofit} and \ref{fig:betafit}, as a proof of principle: average response 
in experiments of the sort performed in Ref.~\cite{geng01} is consistent with an elastic 
continuum approach when microstructure and material order are properly 
incorporated. The results we present are phenomonological in that we 
have obtained elastic coefficients and multipole strengths by fitting 
to data.  We expect that the elastic coefficients we fit are material properties 
in the sense that they could be determined by experiment or simulation 
in another geometry (e.g.~a uniform shear or compression), then used in 
our calculations for point response.

\subsection{Fitting to pressure}
The photoelastic measurements of Geng et al.~associate a scalar quantity 
with each point in space. The measurement technique extracts no 
directional information, so the relevant theoretical prediction to compare to 
experiment is the local pressure $p=(1/2)\Tr\sig$ \cite{gengthesis}.

The data of Ref.~\cite{geng01} are averaged over many experimental realizations; 
the average hydrostatic head is also subtracted. The hydrostatic
contribution to the stress is largest at depth where, as seen below,
the linear (monopole) response dominates.  Therefore, although the
elasticity theory is nonlinear and superposition does not strictly
hold, we expect the incurred error from differencing to be small.  We
note also that our fits necessarily produce regions of small tensile
stress near the surface.  Removal of all tensile stresses from the
solution would require treating the nonlinearity associated with
contact breaking to all orders in the nonlinear elasticity theory. 
In the present context, such regions should be taken only as indicating
that contacts are likely to break.

Fitting to the Cauchy pressure $p$, which is a natural function of the 
deformed coordinates $\x$, presents a difficulty. 
Namely, our caluclations yield a relation 
$\x = \X + \U(\X)$ that is not invertible.  Although in principle 
$\sig$ is known for all points in the deformed material, we can still 
only reference those points by their undeformed positions. That is, we 
have calculated $p(\x(\X))$. Thus for the 
purposes of fitting, we neglect the difference betwen $\x$ and $\X$.
In the experiment, the forcing was restricted to strengths for which the strains were 
small; there were no large-scale rearrangements. This suggests that replacing the 
deformed coordinates with the undeformed coordinates will introduce only small errors. 
Of course, if the strains are small, it is reasonable to ask whether nonlinear elasticity 
is really needed or helpful.  A discussion of this point is
provided in Section~\ref{sec:strainmagnitude} below. 

To facilitate comparison between various materials, we restrict our consideration to boundary forces 
$f = \Qm ka$ with $\Qm = 1$. We have found that similar response profiles can be obtained 
for $0.25 \leq \Qm \leq 2$, and all best-fit values for $\Qm$ lie in this range. 
The force $f=ka$ is that required to compress one Hookean spring through one lattice constant. 

Rather than compare pressure directly to the data of Ref.~\cite{geng01}, we scale each data point by its depth $Z$ 
and  fit to $Z\, P(X,Z)$ for two depths: $Z = 2.7$ cm and 3.60 cm (recall that the grain diameter is 0.80 cm). Scaling by $Z$ compensates for the decay 
of the response with depth. For a reasonable fit, fitting to data at one or two shallow depths gives
good agreement with all data at greater depth. Generally the fitting algorithm returns 
parameters such that agreement with experimental profiles at depths 
shallower than the shallowest fitting depth is poor. 
For the best model material, however, it
is possible to achieve reasonable agreement with data at a depth of 2.25 cm.

\subsection{Pentagonal particles}
\begin{figure}[tbp]
\centering
\includegraphics[clip,width=0.97\linewidth]{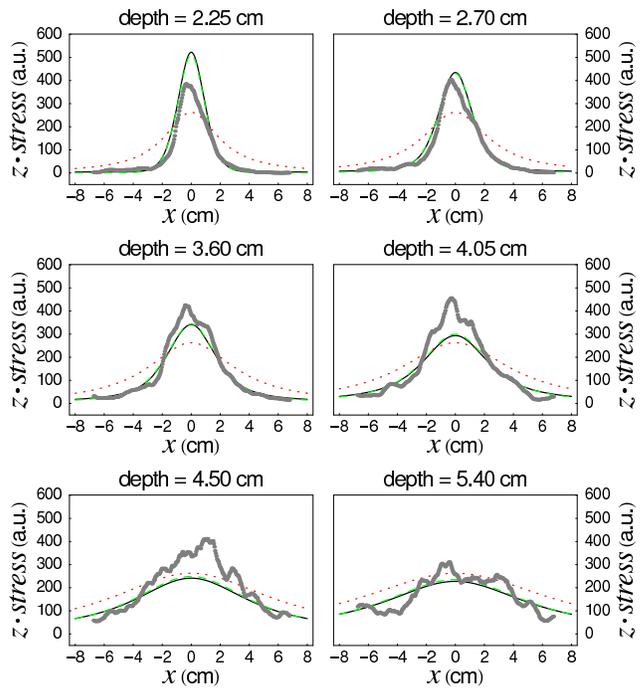}\\
\caption{
(color online) 
(black curves) A fit of Cauchy pressure for a spring-like isotropic ($\beta=0$) material with free 
energy expanded to quartic order in the strains. The fit parameters are $\Qm=1$, 
$\Qd=0.5$, $\Qq=-4.6$, and $k=702$ and were determined by fitting to response in a 
 packing of pentagonal particles (gray points) of width $0.8\,$cm at depths $Z=2.7\,$cm and 3.60\,cm. (dashed green curves) Linear elastic multipole 
response with $\Qm=1$, $\Qd=0.6$,  $\Qq=-4.0$, and $k=700$, fit to the same data.
(dotted red curves) Linear elastic  monopole response with $\Qm = 1$ and $k=1032$. }
\label{fig:isofit}
\end{figure}

The nominal pressure of the spring-like isotropic ($\beta=0$) material for $\Qm=1$, 
$\Qd=0.5$, $\Qq=-4.6$, and $k=702$ is shown in Fig.~\ref{fig:isofit}. 
Parameters were determined by fitting to mean pentagonal particle response data. 
The result is a clear improvement over the fit to linear elastic pressure; 
the nonlinear calculation is able to capture the 
 narrowing of the response as $Z\rightarrow0$. At $Z=2.25\,$cm, shallower than 
 the fitting data, the curve has an appropriate width but overshoots the peak. Note that 
there is little reason {\it a priori} to assume the elastic coefficients we have chosen are the appropriate ones to describe this material. 

A multipole expansion
\begin{equation}
p=\frac{\Qm k a }{\pi R}\cos{\Phi} + \frac{4 \Qd k a^2}{\pi R^2}\cos{2\Phi}
-\frac{2 \Qq k a^3}{\pi R^3}\cos{3\Phi}
\label{eqn:pressuremicro}
\end{equation}
with $\Qm=1$, $\Qd=0.6$, $\Qq=-4.0$, and $k=700$ is nearly indistinguishable 
from the full nonlinear expression with microstructure correction. This 
suggests that in the disordered packings the deviation from monopole-like linear 
elastic response is a consequence of microstructure, not effects captured by the 
nonlinear theory.

\subsection{Hexagonal packings}
\subsubsection{Ball-and-spring fit}
\begin{figure}[tbp]
\centering
\includegraphics[clip,width=0.97\linewidth]{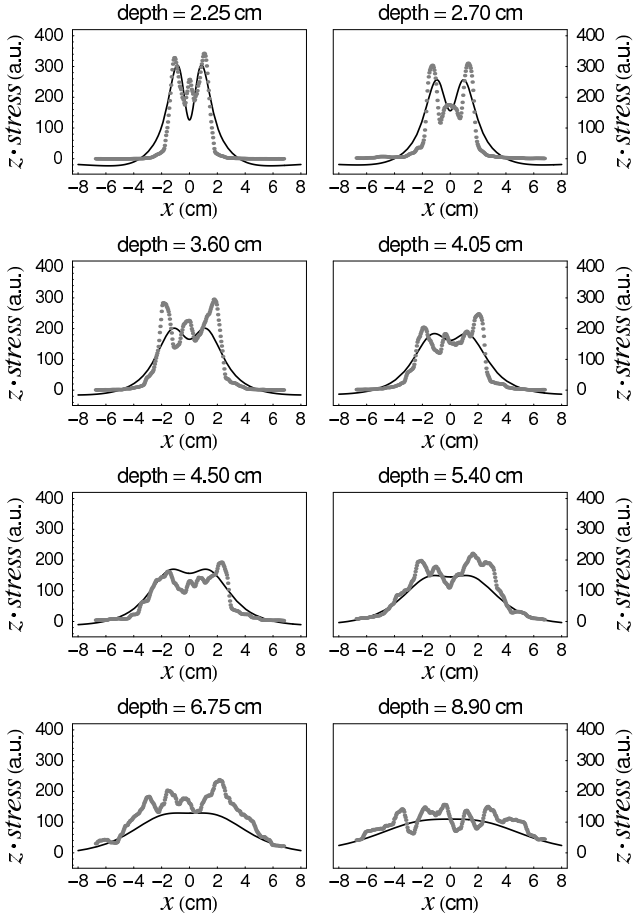}\\
\caption{(black curves)A fit of Cauchy pressure for a ball-and-spring network including cubic and 
quartic corrections to the free energy. The fit parameters are $\Qm=1$, 
$\Qd=9.1$, $\Qq=32$, and $k=112$ and were determined by fitting to response  in a 
monodisperse hexagonal packing of disks (gray points) of diameter $0.8\,$cm at depths 
$Z=2.7\,$cm and 3.60\,cm.   }
\label{fig:springfit}
\end{figure}

The nominal pressure of the ball-and-spring network for $\Qm=1$, $\Qd=9.1$, $\Qq=36$ 
and $k=112$ is shown in Fig.~\ref{fig:springfit}. Parameters were determined by 
fitting to mean hexagonal packing response data. The pressure has 
two peaks at shallow depths; by $Z=5\,$cm it has crossed over to a single central peak. 
As expected, the elastic prediction improves with depth, as the monopole term, which is 
independent of all elastic coefficients, comes to dominate. For depths $z\lesssim 3\,$cm 
there are clear qualitative differences between the fit and the data. The two large peaks in 
the data are wider apart than the prediction and they fall off more sharply with horizontal 
distance from the center; moreover, the theoretical prediction fails to capture the 
small central peak in the data.

\subsubsection{$\alpha$-material fit}
\begin{figure}[tbp]
\centering
\includegraphics[clip,width=0.97\linewidth]{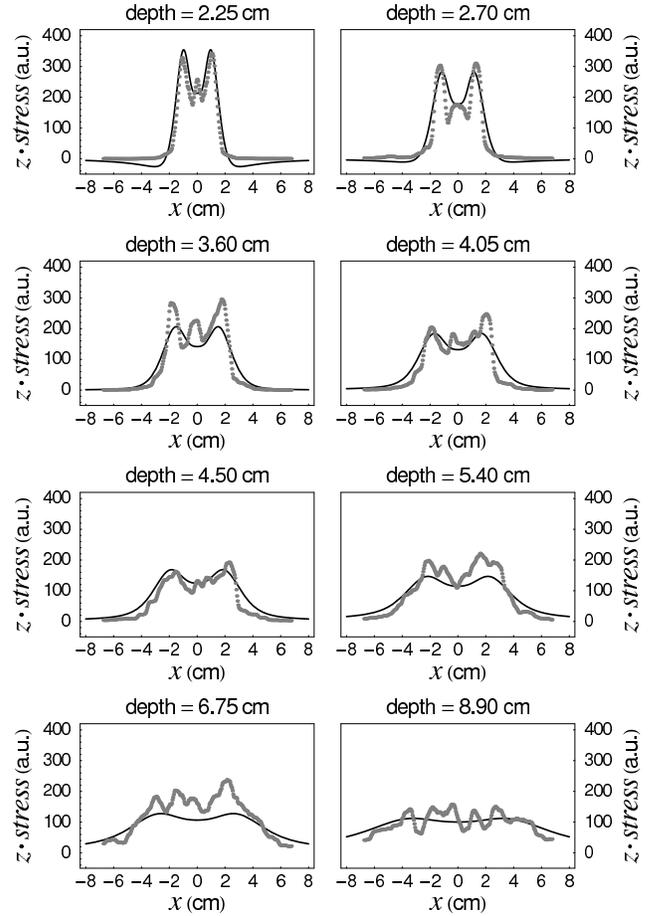}\\
\caption{(black curves) A fit of Cauchy pressure for the $\alpha$-material 
including cubic and quartic corrections to the free energy. 
The fit parameters are $\Qm=1$, 
$\Qd=0.9$, $\Qq=-15$, $k=354$, and $\alpha=8.9$ and were determined by fitting to 
response in a monodisperse hexagonal packing of disks (gray points) 
of diameter $0.8\,$cm at depths $Z=2.7\,$cm and 3.60\,cm.  }
\label{fig:alphafitweak}
\end{figure}

The nominal pressure of the $\alpha$-material for $\Qm=1$, $\Qd=0.9$, $\Qq=-15.4$, 
$k=354$ and $\alpha = 8.9$ is shown in Fig.\ref{fig:alphafitweak}. 
The pressure response in the $\alpha$-material is a better fit than that for the ball-and-spring 
network, as it more closely recreates the two-peaked structure from $Z\approx4\,$cm to 6\,cm. 
It also drops off more sharply in the wings than the ball-and-spring response. The central 
peak, however, is still absent. Moreover, a value of $\alpha\approx 9$ is fairly large 
(see Fig.~\ref{fig:alphaforcelaw}). 

\subsubsection{$\beta$-material fit}
\begin{figure}[tbp]
\centering
\includegraphics[clip,width=0.97\linewidth]{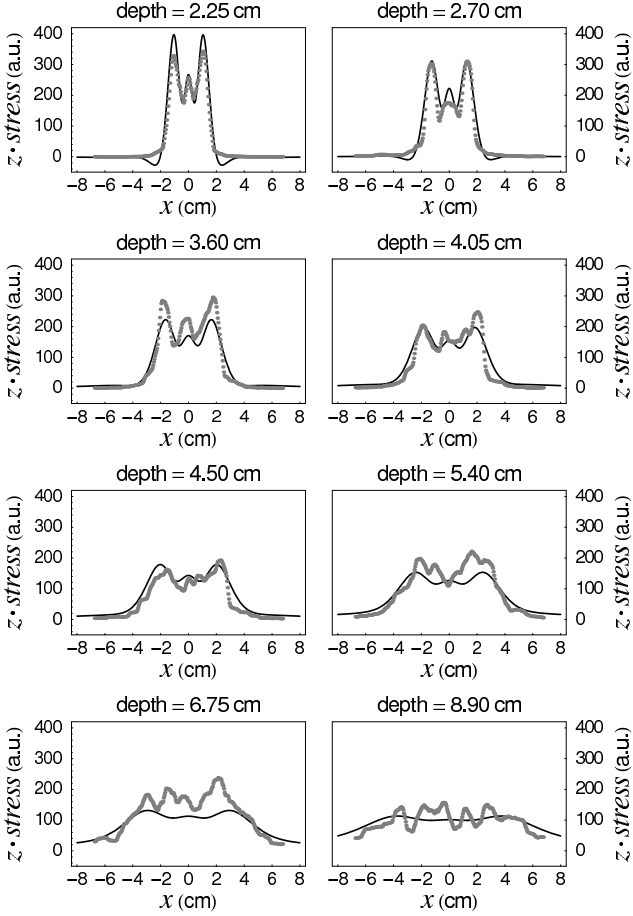}\\
\caption{(black curves) 
A fit of Cauchy pressure for the $\beta$-material including cubic and 
quartic corrections to the free energy. The fit parameters are $\Qm=1$, 
$\Qd=0.6$, $\Qq=-2.0$, $k=353$, and $\beta=12.4$ and were determined by fitting to 
response in a monodisperse hexagonal packing of disks (gray points) of diameter 
$0.8\,$cm at depths $Z=2.7\,$cm and 3.60\,cm.}
\label{fig:betafit}
\end{figure}

The nominal pressure of the $\beta$-material for $\Qm=1$, $\Qd=0.6$, $\Qq=-2.0$, 
$k=353$ and $\beta=12.4$ is shown in Fig.~\ref{fig:betafit}. Parameters 
were determined by fitting to mean hexagonal response data. The 
$\beta$-material response does a better job of capturing the peaks than both the 
ball-and-spring material and $\alpha$-material response profiles. 
Like the $\alpha$-material, the shape of the response peaks of the 
$\beta$-material is narrower and more appropriately positioned 
than that of the ball-and-spring material. The $\beta$-material profiles do a better job 
capturing the small central peak, though the required $\beta$ value of  
$12.4$ represents a hexagonal anisotropy that
is very strong compared to that of a simple ball-and-spring network.

\begin{figure}[tbp]
\centering
\includegraphics[clip,width=0.97\linewidth]{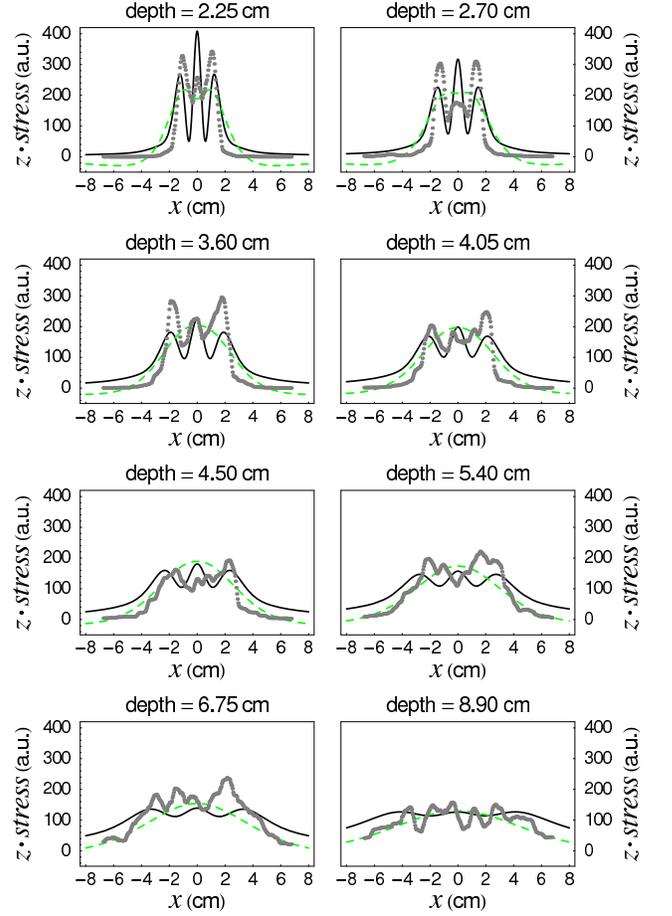}\\
\caption{(color online) 
(black curves) 
A fit of Cauchy pressure for the $\beta$-material including cubic and 
quartic corrections to the free energy but without multipole corrections for 
microstructure ($\Qd = 0 = \Qm$). The fit parameters are $\Qm=1$, $k=509$, and 
$\beta = 10.8$ and were determined by fitting to 
response in a monodisperse hexagonal packing of disks (gray points) of diameter 
$0.8\,$cm at depths $Z=2.7\,$cm and 3.60\,cm. (dashed green curves) 
Linear elastic multipole 
response with $\Qm=1$, $\Qd=11.4$,  $\Qq=43$, and $k=116$, fit to the same data. }
\label{fig:betafit2}
\end{figure}

Fig.~\ref{fig:betafit2} shows the $\beta$-material response without microstructure 
correction ($\Qm=1$, $\beta = 10.8$, $k=509$) and the linear elastic response with 
induced multipole terms of Eq.~(\ref{eqn:pressuremicro}) 
($\Qm=1$, $\Qd=11.4$, $\Qq=42$, $k=116$). Neither agrees well with the data. 
It is necessary to include nonlinear as well as microstructure corrections
to the linear elastic result to obtain good agreement with the mean hexagonal response 
data. This contrasts with the mean disordered response data, which can be described 
with a microstructure correction alone. We infer that nonlinear corrections are needed 
in the hexagonal system to capture the material anisotropy.

\subsection{Crossover to linear elasticity}
\begin{figure}[tbp]
\centering
\includegraphics[clip,width=0.5\linewidth]{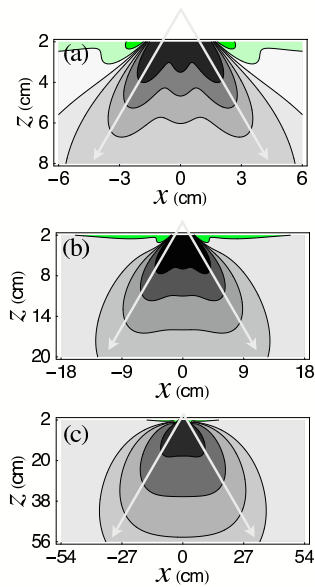}
\caption{(color on-line) Pressure contours for the $\beta$-material with fit parameters identical to 
those in Fig.~\ref{fig:betafit}.
At shallow depths the structure is three-peaked, the outer two seeming to propagate with 
depth. At greater depth the crossover to monopole response is evident. Regions of tensile 
stress near the surface are plotted in green.}
\label{fig:hex_contours}
\end{figure}

For shallow depths the hexagonal anisotropy of the ordered disk packing is strongly reflected 
in the functional form of its stress response. The dipole and quadrupole corrections which 
shape the response in the near field fall off as $1/R^2$ and $1/R^3$, respectively, while the 
monopole response decays as $1/R$. Sufficiently deep within 
the material, the monopole term, which is identical to the linear elastic solution, will dominate. 
Fig.~\ref{fig:hex_contours} shows contours of the nominal pressure for the $\beta$-material of Fig.~\ref{fig:betafit} in the near and 
far fields. In the first $6\,$cm of depth the three peaks seen in the data are clearly visible.
The contours of the pressure in linear elasticity are circles, and by a depth of 
$40\,$cm this form is largely recovered.

\subsection{Physical pressure and strain}
\begin{figure}[tbp]
\centering
(a)\includegraphics[clip,width=0.65\linewidth]{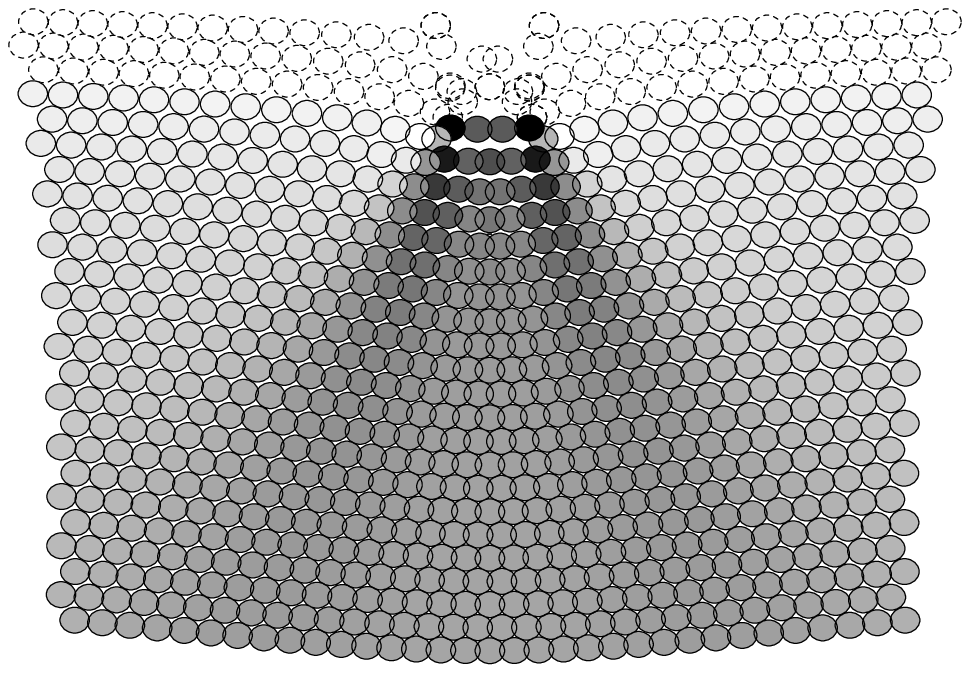}\\
(b)\includegraphics[clip,width=0.65\linewidth]{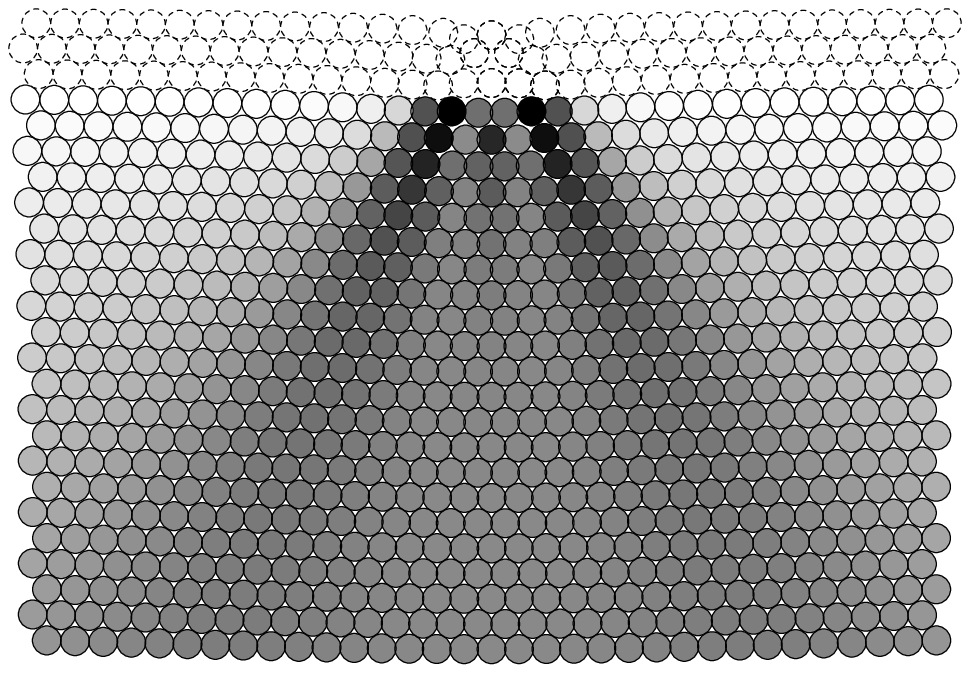}\\
\caption{(a) The deformed $\beta$-material. The first three layers of the material are omitted. 
Disk $i$ with lattice position
$\X_i$ in the undeformed material is shown here centered at $\x_i$. Each disk is shaded 
according to $R_i p_i$, the physical pressure scaled by the (undeformed) distance from the 
point force; values increase from blue through purple and orange to green. Pressures are 
calculated for the case 
for $\Qm=1$, $\Qd=0.6$, $\Qq=-2.0$, $k=353$ and $\beta=12.4$. Two-peaked structure is apparent, 
as well as arching in the upper layers. The strains are large. (b) The deformed $\beta$-material for $\Qm=1/4$, $\Qd=0.12$, $\Qq=-1.2$, $k=1415$ and $\beta=45.7$. }
\label{fig:betapacking}
\end{figure}

Having determined fit parameters, it is possible to visualize the physical or Cauchy pressure 
$p=(1/2)\mathrm{Tr}\sig(\x(\X))$ and 
strains in the material. In the undeformed material, each disk sits on a lattice site which we 
label by an index $i$. Under the deformation the disk at $\X_i$ moves to 
$\x_i = \X_i + \U_i$. We draw a disk of radius $D=0.8\,$cm at $\x_i$ and shade it in 
proportion to $|\X_i| p_i(\x_i(\X_i))$. The first three layers of the packing, for which 
the displacements and pressure are clearly diverging near the applied force, are drawn but not shaded. 
Though we do not make any attempt to portray the deformations of the disks themselves, the  
overlap or separation between disks gives a good sense of the strain in the material, and the colormap 
indicates the local variation of pressure on the grain scale. The $\beta$-material fit for 
$\Qm=1$ is shown in Fig.~\ref{fig:betapacking}. The two-peaked response structure is immediately 
apparent; the smaller third peak is more difficult to see, but present for the first few rows. 
There is dilatation near the surface. The disks directly 
below the applied force participate in the formation of arches, which is consistent with the 
appearance of two large peaks along the lines $\phi = \pm \pi/6$.

\section{Strain magnitude} \label{sec:strainmagnitude}

We have demonstrated that hexagonally anisotropic nonlinear elastic
response can display stress profiles similar to those seen in ordered
granular packings, which suggests that significant deviations from the 
classical Boussinesq response can extend to
depths of tens of layers. However, from Fig.~\ref{fig:betapacking}a it
is also clear that the attendant strains are large, creating regions
of strains in the first two dozen layers that are much larger than
those observed in the systems studied by Geng et al. This is not
entirely surprising for the choice $\Qm=1$. We note, however, that by
fixing $\Qm=1/4$, as in Fig.~\ref{fig:betapacking}b, we obtain a fit
in which strains outside the first three layers are reasonably small. 
Differences from the response profiles in Fig.~\ref{fig:betafit} are
imperceptibly small; plotted on top of Fig.~\ref{fig:betafit}, the 
$\Qm=1/4$ and $\Qm=1$ curves would overlap.
The microstructure corrections are still of order
unity, the spring constant is four times larger (so that the imposed
force $f$ is unchanged), and the hexagonal anisotropy is increased
significantly: $\beta = 45.7$.  Thus in a our simplistic
ball-and-spring-inspired material, the observed profiles can be
attributed either to strong nonlinearity due to large strain magnitude
or to strong hexagonal anisotropy.

The material constants of Eq.~(\ref{eqn:betamaterial}) were chosen as
a minimal hexagonally anisotropic model, rather than derived from
a microscopic model.  
We speculate that the enhancement of the nonlinearity and/or the hexagonal
anisotropy over the values obtained naturally from simple ball-and-spring
models may be due to the importance of a short length scale
$\delta \ll D$ in the grain-grain interactions.
Such a length scale may
be the consequence of, e.g., nonlinear grain interactions (``soft
shell'' grains \cite{degennes99} or Hertzian force laws), or
inhomogeneous elastic coefficients due to microscopic grain
irregularities \cite{gg02, didonna05}, in which case small strains
may correspond to large deformations of contacts on the relevant scale $\delta$.
Full consideration of such effects is beyond the scope of the present
work.

Considering all the results presented above, we arrive at the following picture. 
The important distinctions between 2D disordered and hexagonal granular packings 
are the effects near the applied point force and the material symmetry. 
Although nonlinearity complicates calculations considerably, it enters only as a 
matter of necessity in incorporating material order: elasticity cannot distinguish 
isotropic and hexagonally anisotropic materials otherwise. The facts that 
1) nonlinearities  in the isotropic material provide no notable improvement over 
microstructure corrections alone (see Fig.~\ref{fig:isofit}), and 2) hexagonal 
materials admit reasonable response profiles for small strain and strong anisotropy 
(see Fig.~\ref{fig:betapacking}b), underscore this point. A large 
$\beta$ value may be difficult to interpret in terms of a microscopic model, but 
this is not surprising given that it represents a combination of strong local 
nonlinearites and an ensemble average over microstructures that are known to lead 
to vastly different stress or force chain patterns.

\section{Conclusion}
Our results indicate that continuum elasticity theory can provide
semi-quantitative explanations of nontrivial experimental results on
granular materials.  For isotropic (disordered) materials subject to a
point force, it appears that nonlinearities are less important than
multipoles induced at the surface where continuum theory breaks down.
For hexagonal disk packings, however, the anisotropy associated with
nonlinear terms in the elasticity theory is required.  We have studied
the nonlinear theory of response in a hexagonal lattice of point
masses connected by springs and a phenomenological free energy with an
adjustable parameter determining the strength of the hexagonal
anisotropy.  A similar treatment would be possible for systems with, 
e.g.~, square or uniaxial symmetry, but the free energy would acquire 
additional terms at all orders. For a particular choice of elastic 
coefficients, the multiple peaks in the pressure profile at intermediate 
depths and the recovery of the familiar single peak of conventional 
(linear) elasticity theory at large depths are well described by the 
theory. To the extent that theoretical approaches based on properties of
isostatic systems predict hyperbolic response profiles \cite{blumenfeld04}, our
analysis indicates that the materials studied in Refs.~\cite{geng01} and \cite{geng03}
have average coordination numbers that place them in the elastic
rather than isostatic regime.

\acknowledgments
We thank R.~Behringer and J.~Geng for sharing their data with us.  We also thank
D.~Schaeffer and I.~Goldhirsch for useful conversations.  This work was supported
by the National Science Foundation through Grant NSF-DMR-0137119. BT acknowledges 
support from the physics foundation FOM for portions of this work done in Leiden.

\appendix
\section{Pressure in the $\beta$-material}
The expression of Eq.~(\ref{eqn:pressure}) gives the pressure in a 
horizontally oriented ball-and-spring network with elastic coefficients 
given by Eq.~(\ref{eqn:hexagonalenergy}). The pressure in the 
$\beta$-material of Eq.~(\ref{eqn:betamaterial}) is given by adding 
an additional term $p_\beta(r(R,\Phi), \phi(R,\Phi))$ to the expression 
in Eq.~(\ref{eqn:pressure}), where $p_\beta$ is given by
\begin{widetext}
\begin{eqnarray}
p_\beta(r(R,\Phi),\phi(R,\Phi)) 
&=& \frac{B_2}{R^2}\beta \biggl[ -6c_4 - 9c_6 - 4c_8 \biggr] \nonumber \\
&& + \frac{B_3}{R^3}\beta \biggl[
\frac{28333}{420}c_3 + \frac{205}{2}c_5 + \frac{139}{3}c_7 + 
\frac{35}{2}c_9 \biggr. \nonumber \\
&& \hphantom{\frac{B_3}{R^3}\beta\biggl[}
+2\Phi \biggl(-62 s_3 +33 s_5 + 44s_7 + \frac{55}{3}s_9 \biggr) \biggr.
% \nonumber \\
% && \hphantom{\frac{B_3}{R^3}\beta\biggl[}
+L\biggl(-50c_3 + 36c_7 + 30c_9 \biggr) \biggr] \nonumber \\
&& + \frac{B_3}{R^3}\beta^2 \biggl[
9c_1 - \frac{10418}{735}c_3 + \frac{15}{2}c_9 + \frac{63}{4}c_{11} \biggr.
% \nonumber \\
% && \hphantom{\frac{B_3}{R^3}\beta^2\biggl[}
+\frac{119}{10}c_{13} + \frac{10}{3}c_{15} + 36\Phi s_3 + 24L c_3 \biggr]
\nonumber \\
&& + \frac{B_3^\prime \Qd}{\sqrt{3}\pi R^3}\beta \biggl[
\frac{5\sqrt{3}\pi}{14}c_3 + \frac{28333\pi}{6720}c_3 
+ \frac{205\pi}{32}c_5 -6\sqrt{3}\pi c_7
+\frac{139}{48}c_7 - 5\sqrt{3}\pi c_9 + \frac{35}{32} c_9 \biggr.
\nonumber \\
&& \hphantom{+ \frac{B_3^\prime\Qd}{\sqrt{3}\pi R^3}\beta \biggl[}
+\Phi \biggl(-\frac{31}{4}s_3 + \frac{33}{8}s_5 + \frac{11}{2}s_7 
-\frac{55}{24}s_9 \biggr) 
+L\biggl( -\frac{25}{8}c_3 + \frac{9}{4}c_7 + \frac{15}{8}c_9 \biggr)
\biggr] \nonumber \\
&& + \frac{B_3^\prime \Qd}{\sqrt{3}\pi R^3} \beta^2 \biggl[
\frac{9}{16}c_1 - \frac{5209}{5880}c_3 + \frac{15}{32}c_9 
+ \frac{63}{64}c_{11} + \frac{119}{160}c_{13} 
+\frac{5}{24}c_{15} \biggr. 
%\nonumber \\
%&& \hphantom{+ \frac{B_3^\prime \Qd}{\sqrt{3}\pi R^3} \beta^2 \biggl[ }
+\frac{9}{4}\Phi s_3 + \frac{3}{2}L c_3 \biggr]
\end{eqnarray}
\end{widetext}

% \bibliographystyle{apsrev}
% \bibliography{thesisbib}

\end{document}